\documentstyle[epsfig,aps,floats,preprint]{revtex}

\newcommand{\beq}{\begin{equation}}
\newcommand{\eeq}{\end{equation}}
\newcommand{\bea}{\begin{eqnarray}}
\newcommand{\eea}{\end{eqnarray}}

\def\laq{\raise 0.4ex\hbox{$<$}\kern -0.8em\lower 0.62
ex\hbox{$\sim$}}
\def\gaq{\raise 0.4ex\hbox{$>$}\kern -0.7em\lower 0.62
ex\hbox{$\sim$}}

\def \ra {\rightarrow}

\def \b {\beta}
\def \a {\alpha}

\def \Ga {\Gamma}
\def \ga {\gamma}
\def \sg {\sigma}
\def \da {\delta}
\def \ep {\epsilon}
\def \r {\rho}
\def \om {\omega}

\begin{document}
\par
\begingroup

\begin{flushright}
DFTT-23/98\\
gr-qc/9805060\\
\end{flushright}

\vspace{10mm}
{\large\bf\centering\ignorespaces
Repulsive gravity in the very early Universe
\vskip2.5pt}
{\dimen0=-\prevdepth \advance\dimen0 by23pt
\nointerlineskip \rm\centering
\vrule height\dimen0 width0pt\relax\ignorespaces
 M. Gasperini
\par}
{\small\it\centering\ignorespaces
Dipartimento di Fisica Teorica, Universit\`a di Torino, 
Via P. Giuria 1, 10125 Turin, Italy \\
and Istituto Nazionale di Fisica Nucleare, Sezione di Torino, 
Turin, Italy \\
\par}
{\small\rm\centering(\ignorespaces March 1998\unskip)\par}

\par
\bgroup
\leftskip=0.10753\textwidth \rightskip\leftskip
\dimen0=-\prevdepth \advance\dimen0 by17.5pt \nointerlineskip
\small\vrule width 0pt height\dimen0 \relax

\begin{abstract}
I present two examples in which the curvature singularity of a
radiation-dominated Universe is regularized by {\sl (a)} the repulsive
effects of spin interactions, and {\sl (b)} the repulsive effects arising
from a breaking of the local gravitational gauge symmetry.
In both cases the collapse of an initial, asymptotically flat state is
stopped, and the Universe bounces towards a state of decelerated
expansion. The emerging picture is typical of the pre-big bang
scenario, with the main difference that the string cosmology dilaton
is replaced by a classical radiation fluid, and the solutions are not
duality-invariant.
\end{abstract}

\vspace{10mm}
\begin{center}
---------------------------------------------\\
\vspace {5 mm}
{\sl Awarded the ``Fourth Prize" in the 1998 Awards
for Essays on Gravitation} \\
{\sl (Gravity Research Foundation, Wellesley Hills, MA)}\\
\bigskip
To appear in {\bf Gen. Rel. Grav.}
\end{center}
\vspace{5mm}

\par\egroup
\thispagestyle{plain}
\endgroup

\pacs{}

The aim of this Essay is to discuss the possibility of avoiding the
initial cosmological singularity through a phase of repulsive gravity
occurring in the very early Universe. I will consider two mechanisms
of repulsive gravity: spin-torsion interactions and spontaneous
breaking of the local $SO(3,1)$ gauge symmetry. I will show that in
both cases the condition of geodesic convergence \cite{1} can be 
violated, and the cosmological 
equations may admit regular homogeneous and 
isotropic solutions for which the energy density and the curvature
grow up to a maximum (finite) scale, and then decrease, with a
smooth joining to the standard decelerated evolution.

The interesting aspect of such models is that they do not require any
violation of the strong energy condition \cite{1} in the
conventional matter sector. Indeed, in both cases I will simply take a
radiation-like equation of state for the sources (no vacuum energy
term will be included). In spite of the fact that I will use classical
generalization of the Einstein equations, the results obtained might
be of some relevance for applications to string cosmology, where the
present cosmological phase is expected to emerge from a phase of
growing curvature, through a smooth transition that should avoid the
initial singularity \cite{2}. 

I will first discuss the case of spin-torsion interactions.
Torsion is a natural ingredient of gauge theories of the Poincar\`e
group \cite{3}, as it represents the field strength of local
translations, and it is thus the required Yang-Mills partner of the
curvature (the field strength of local Lorentz rotations). In addition,
torsion couples minimally to the axial current of spinor matter, as
required by local supersymmetry: simple supergravity, containing
only the graviton and the gravitino, can indeed be formulated as an
Einstein-Cartan theory for the Rarita-Schwinger field \cite{4}. 

The Einstein-Cartan theory \cite{3}, which I will consider in this
paper, is the simplest example of gravitational theory with torsion.
In such a theory torsion does not propagate, and it can be
non-vanishing only in the presence of an intrinsic spin density of
matter. As a consequence, no significant effect is expected for
macroscopic bodies at ordinary densities; torsion interactions may
become important, however, in the regime of extremely high density
and curvature of the early Universe.

Let us thus consider a cosmological application of the
Einstein-Cartan theory, by taking a perfect gas of spinning particles
as the effective matter source. In that case the connection is
non-symmetric, $\Ga_{[\mu\nu]}{}^\a \not= 0$, and besides the
equation relating the Einstein tensor and the canonical
(non-symmetric) energy-momentum tensor, 
\beq
G_{\mu\nu}(\Ga)= 8\pi G T_{\mu\nu},
\label{1}
\eeq
we have an additional algebraic relation \cite{3} between
the torsion,  $Q_{\mu\nu}{}^\a=\Ga_{[\mu\nu]}{}^\a$, and the
canonical spin density tensor, $S_{\mu\nu}{}^\a$:
\beq
Q_{\mu\nu}{}^\a= 8 \pi G \left(S_{\mu\nu}{}^\a +{1\over
2}\da_\mu^\a S_{\nu\b}{}^\b-{1\over
2}\da_\nu^\a S_{\mu\b}{}^\b\right). 
\label{2}
\eeq
Thanks to the above relation, torsion can be eliminated everywhere
in eq. (\ref{1}). By assuming a convective model of
spinning fluid minimally coupled to the geometry of the
Riemann-Cartan manifold \cite{5}, we can rewrite eq. (\ref{1}) in the
standard Einsteinian form for a symmetric connection, but with
additional terms that are linear and quadratic in the spin tensor of
the matter sources.  

In the absence of some  external polarizing field the spins are
randomly oriented, and the linear terms are zero after an
appropriate space-time averaging, $\langle S_{\mu\nu\a}\rangle
=0$; the quadratic terms, however, are non-vanishing also on the
average,  $\langle S_{\mu\nu\a}S^{\mu\nu\a}\rangle \not=0$.
Because of the spinning sources we are thus led to a modified set of
cosmological equations, even for unpolarized matter, and in the
averaged macroscopic limit. For a spatially flat metric
$g_{\mu\nu}=$ diag $ (1, -a^2 \da_{ij})$, in particular, the
averaged cosmological equations can be written as \cite{6}:
\bea
&&
H^2={8\pi G \over 3} \left(\r-2\pi G \sg^2\right) , 
\label{3}\\
&&
\dot H +H^2=-{4\pi G \over 3} \left(\r+3 p -8\pi G \sg^2\right) .
\label{4}
\eea
Their combination gives the conservation equation 
\beq
\dot \r -2\pi G (\sg^2)\dot{} +3H\left(\r+p-4\pi G \sg^2\right)=0,
\label{5}
\eeq
where $H=\dot a/a$, and a dot denotes differentiation with respect
to cosmic time. I have defined $\sg^2=
\langle S_{\mu\nu\a}S^{\mu\nu\a}\rangle/2$, and $\r$, $p>0$ are the
energy density and the pressure of the fluid in the zero spin limit. 

When $8\pi G \sg^2> \r+3 p$ the condition of geodesic convergence is
violated,
\beq
R_{\mu\nu}u^\mu u^\nu= -3 \left(\dot H +H^2\right) <0,
\label{6}
\eeq
even if the pressure satisfy the strong energy condition, $\r+3p>0$.
In a previous paper this repulsive contribution of the spin density
was used to discuss the possibility of spin-dominated inflation
\cite{6}. Here it will be used for a possible regularization of the
initial curvature singularity. 

The spin contribution to the geometry depends, of course, on the
particular model of fluid. In order to show that this repulsive
interaction can be strong enough to allow a smooth cosmological
evolution, I will consider a spinning liquid of
unpolarized fermions \cite{7}, with equation of state $p=\ga \r$,
and averaged  squared spin tensor $\sg^2 \propto \r^{2/(1+\ga)}$. In
this case the equations can be integrated exactly. For relativistic
fermions, in particular, we have $\ga=1/3$ , the conservation
equation (\ref{5}) gives $\r \propto a^{-4}$, and the integration of
eq. (\ref{3}) leads to
\beq
 {t\over l_p}\sqrt{8 \pi \over 3}=
{a\over 2} \sqrt{c_1 a^2 -c_2} + {c_2\over 2} \ln \left|a+
 \sqrt{c_1 a^2 -c_2}\right|
\label{7}
\eeq
($c_1, c_2$ are dimensionless positive constants, and we are
measuring time in Planck length units, with $\l_p=\sqrt G$). 

A plot of the energy density and of the Hubble parameter for this
solution is shown in Fig. 1. The curvature is everywhere regular, and
the models describes a smooth evolution from a phase of
accelerated contraction, growing curvature, to a phase of
decelerated expansion, decreasing curvature. The scale factor 
contracts down to a minimal value $a_m= \sqrt{c_2/c_1}$, and then
re-expands (like $a \sim t^{1/2}$, asymptotically).  In string
cosmology, this behaviour is typical of the pre-big bang scenario 
represented in terms of the Einstein frame metric \cite{8}. 

\begin{figure}[t]
\begin{center}
\mbox{\epsfig{file=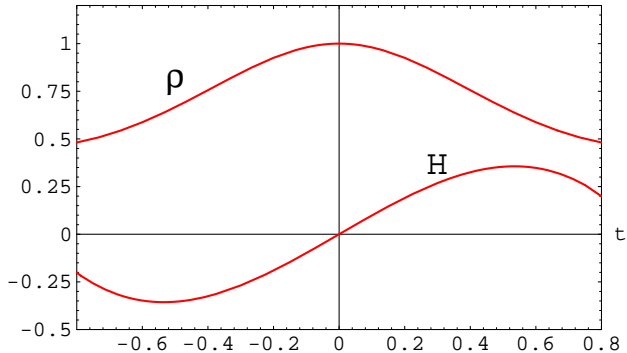,width=82mm}}
\vskip 5mm
\caption{\sl Time evolution of the Hubble factor and of the energy
density according to eq.  (\ref{7}). I have put $c_1=c_2=1$, and time 
is measured  in units of $(3/8\pi)^{1/2}l_p$.} 
\end{center}
\end{figure}

It may be interesting to observe that a similar class of solutions can
also be obtained from the string cosmology equations through a
duality boost of the flat, two-dimensional Milne metric \cite{9}. 
Indeed, this fact is more than a coincidence, as the global $O(3,3)$
duality group, used in \cite{9}, introduces a non-trivial
antisymmetric tensor background, $H_{[\mu\nu\a]} \not= 0$, which is
known to have a geometric interpretation as the torsion of an
appropriate connection. The main difference is that in string
cosmology the ``matter source" is the scalar
dilaton field, while in this example matter is more conventionally
represented as a perfect fluid, and the duality symmetry of string
theory is lost. 

A second, possible mechanism for the generation of repulsive
interactions in the early Universe is associated to the breaking of the
local $SO(3,1)$ symmetry of the gravitational interaction \cite{10}. 
This symmetry is part of the local gauge group of gravity: in the 
gauge approach to general relativity, the anholonomic Ricci
connection $\om_\mu ^{i j}$ represents in fact the Yang-Mills
potential of local Lorentz rotations, which transforms as a covariant
vector in the index $\mu$ under general reparametrizations, and as
an antisymmetric tensor in the two ``internal" indices $i,j$, under
the action of the local $SO(3,1)$ group. 

Like every gauge symmetry, also this local Lorentz symmetry can be
broken spontaneously when an appropriate (geometric) potential,
generated by a  self-interacting antisymmetric tensor,  
appears in the action \cite{11}. This breaking leads to an effective
``quasi-riemannian" theory \cite{12}, namely to a gauge theory of
gravity invariant under general reparametrizations, but with a local
tangent space group other than the Lorentz group. From a
phenomenological point of view, the main consequences of such a
breaking  are the possible appearance of repulsive forces,
\cite{10,11}, and the possible violation of the equivalence principle
\cite{13,14}. 

The violation of the weak equivalence principle, however, is not a
necessary consequence of any Lorentz symmetry breaking. If we
consider, for instance, a four-dimensional quasi-riemannian theory
with local $SO(3)$ invariance, we find that the most general model
contains four independent parameters in the gravitational part of the
action, and three parameters in the matter action. By
imposing four conditions on these seven parameters it is always
possible to preserve the covariant conservation of the energy
momentum tensor, in such a way that the motion of test particles
remains geodesic \cite{13}.  

In that case the causal structure of space-time is still determined by
the metric tensor,  the classical singularity theorems \cite{1} still
can be applied, and the violation of geodesic
convergence is still a necessary condition for singularity prevention.
Because of the modified dynamical equations, however, geodesic
convergence and strong energy condition are no longer equivalent
\cite{15}, so that a smooth and complete model of cosmological
evolution can be implemented even with conventional matter
sources, satisfying the strong energy condition. 

As a particular example of this possibility I will consider here a
one-parameter, $SO(3)$-invariant quasi-riemannian model of gravity,
which for a closed, homogeneous and isotropic manifold is described
by the action
\beq
S= 16 \pi G S_m -\int dt a^3 \left[(1+\ep){6H^2\over N} -{6 k\over
a^2} N\right] .
\label{8}
\eeq
Here $S_m$ is the action for perfect fluid matter, $N$ is the lapse
function, $k$ is the spatial curvature (in Planck length units), and
$\ep$ is a dimensionless constant parametrizing the breaking of the
local Lorentz symmetry.  All the other parameters have been fixed in 
such a way as to 
preserve the geodesic motion of the cosmological fluid \cite{15}. In
the limit $\ep \ra 0$ the action reduces to the standard, general
relativistic action. 

The variation with respect to $N$ and $a$, in the cosmic time gauge
$N=1$, leads to the equations
\bea
&&
(1+\ep) H^2 +{k\over a^2} = {8\pi\over 3} G \r ,
\label{9}\\
&&
(1+\ep)\left(2\dot H +3 H^2\right) +{k\over a^2} = -{8\pi} G p ,
\label{10}
\eea
and their combination gives 
\beq
\dot \r +3H\left(\r+p\right)=0,
\label{11}
\eeq
in agreement with the weak equivalence principle, $\nabla_\nu
T_\mu {}^\nu=0$. Note that in the absence of spatial curvature this
particular breaking of the gauge symmetry has no effect on a
cosmological metric, apart from a trivial renormalization of the
gravitational coupling constant.

The value of $\ep$ depends on the parameters of the antisymmetric
tensor potential \cite{11} that breaks spontaneously $SO(3,1)$ down
to $SO(3)$. Today, and at a macroscopical level,  a breaking  of local
Lorentz symmetry  is strongly constrained by many experimental
data \cite{13,14}.  In the regime of extremely high temperature and
density of the very early Universe, however, such phenomenological
constraints do not necessarily apply, and for $\ep <-1$ 
gravity may become repulsive enough to prevent the singularity,
even if $\r+3p>0$. 

Consider in fact a radiation fluid, $p=\r/3$, so that, 
from eq. (\ref{11}), $\r=\r_0 a^{-4}$. The integration of eq. (\ref{9}),
for $k=+1$ and $\ep <-1$ , gives then 
\beq
a(t)= \left[ {8\pi\over 3} \r_0 \l_p^4 + {1\over |1+\ep|}
\left(t \over \l_p\right)^2 \right]^{1/2},
\label{12}
\eeq
where $\r_0$ is a positive integration constant. For this solution, the
plot of the Hubble parameter
\beq
H= {t\over t^2 +|1+\ep| {8\pi\over 3} \r_0 \l_p^6}
\label{13}
\eeq
and of  the energy density 
is qualitatively the same as the plot of Fig. 1: the
initial collapse of an asymptotically flat state is stopped, and the
Universe bounces to a state of curvature-dominated, linear
expansion. Note however that, unlike the Einstein-Cartan solution of the
previous example, in this case the Universe does not become
asymptotically radiation-dominated.

In conclusion, I would like to stress the fundamental role played by
antisymmetric tensors in these two examples of regular
cosmological models. In the first case the repulsive forces stopping
the collapse are due to the coupling between the spin and the
antisymmetric torsion field, in the second case they are due to a
self-interacting antisymmetric tensor that provides the right ``Higgs
potential" for the breaking of the local  $SO(3,1)$ symmetry. This 
suggests that a successful, singularity-free pre-big bang scenario
might require a non-trivial antisymmetric tensor background, arising
either from the NS (Neveu-Schwartz) or  the RR (Ramond-Ramond)
sector of the underlying string theory (or M-theory) effective action
\cite{16}.

\end{document}